\documentclass[journal]{IEEEtran}
\usepackage[nocompress]{cite}
\usepackage{amsmath,amssymb,amsfonts}
\usepackage{algorithmic,algorithm}
\usepackage{graphicx}
\usepackage{textcomp}
\usepackage{bm}

\usepackage{xcolor}

\def\BibTeX{{\rm B\kern-.05em{\sc i\kern-.025em b}\kern-.08em
    T\kern-.1667em\lower.7ex\hbox{E}\kern-.125emX}}

\begin{document}

\author{Hassan Touati, \IEEEmembership{Member, IEEE},
Rodrigo C. de Lamare, \IEEEmembership{Fellow, IEEE}

\thanks{The authors are with the Department of Electrical Engineering (DEE), Pontifical Catholic University of Rio de Janeiro (PUC-Rio), Rio de Janeiro 22451-900, Brazil}}

\title{Study of Adaptive Reliability-Driven Conditional Innovation Decoding for LDPC Codes}
\maketitle
\begin{abstract}
In this work, we present an adaptive reliability-driven conditional innovation (AR-CID) decoding algorithm for low-density parity check (LDPC) codes. The proposed AR-CID decoding algorithm consists of one stage of message quality checking and another stage of message passing refinement, which are incorporated into a residual belief propagation decoding strategy. An analysis of the AR-CID decoding algorithm is carried out along with a study of its computational complexity and latency characteristics. Simulation results for several examples of LDPC codes, including short and medium-length codes over an extended range of channel conditions, indicate that the proposed AR-CID decoding algorithm outperforms competing decoding techniques and has an extremely fast convergence, making it particularly suitable for low-delay applications.  
\end{abstract}

\section{Introduction}
\label{sec:introduction}
In modern communication networks, reliable information transmission is paramount, underlying systems ranging from internet communication and satellite transmissions to mobile networks. At the core of these systems are Low-Density Parity Check (LDPC) codes \cite{LDPC}, which play a crucial role in ensuring accurate data recovery, even when signals are distorted by interference and noise. LDPC codes have been adopted in numerous standards, including 5G New Radio (NR) \cite{5G_NR}, IEEE 802.11n/ac/ax Wi-Fi standards, and DVB-S2 satellite communications due to their capacity-approaching performance and structural flexibility.

Dynamic scheduling techniques such as Informed Dynamic Scheduling (IDS) \cite{IDS, ids} have accelerated the decoding of LDPC codes by optimizing the sequence and timing of message updates. These strategies improve efficiency and performance, outperforming traditional approaches such as Flooding Belief Propagation (BP) \cite{ BP1}, which simultaneously updates all nodes of the Tanner graphs of LDPC codes but requires more iterations to converge. Layered Belief Propagation (LBP) \cite{LBP, LBP1 ,LBP2025} updates nodes sequentially on a layer-by-layer fashion, achieving faster convergence with improved memory efficiency. Residual Belief Propagation (RBP) \cite{RBP, RBP_1, RBP2} prioritizes updates with the largest residuals for fast decoding, while Residual-Decaying RBP (RD-RBP) \cite{RD_RBP} gradually reduces the magnitude of residuals to refine decoding. 

Reliability-based methods such as the reliability profile (RP) \cite{RP} and Conditional Innovation (CI) \cite{CI} perform selective updates using Log-Likelihood Ratios (LLRs), enhancing both speed and accuracy. Furthermore, Uniformly Reweighted BP (URW) \cite{URW} ensures balanced updates through systematic reweighting, the variable factor appearance probability belief propagation (VFAP-BP) algorithm can further exploit the cycles by performing a selective reweighting \cite{vfap}, knowledged-aided IDS approaches \cite{kaids} and List-RBP \cite{List_RBP} combines prioritization with breadth by updating a ranked list of residuals. Alternative paradigms such as Guessing Random Additive Noise Decoding (GRAND) \cite{GRAND}, which represents a fundamental shift from traditional message-passing by guessing noise patterns rather than iteratively refining beliefs.

In this work, we propose an adaptive reliability-driven conditional innovation (AR-CID) decoding algorithm for LDPC codes \cite{arcid}. The proposed AR-CID decoding algorithm consists of one stage of message quality checking and another stage of message passing refinement, which are incorporated into a residual belief propagation decoding strategy. The message quality checking is inspired by the contextual information transition and the message fidelity index that can inform AR-CID about the quality of the messages prior to further processing. Depending on the assessment of the messages at the first stage, AR-CID employs a modified version of RBP that employs old and new messages in the update. 

Compared to the Conditional Innovation (CI) method \cite{CI}, AR-CID provides several key improvements: (1) \textit{Adaptive Selection}: while CI uses a fixed criterion for node updates, AR-CID dynamically adapts the selection of $\lambda N$ nodes based on both reliability index $R_v$ and contextual information transition $\Delta_y$. (2) \textit{Two-stage Assessment}: AR-CID incorporates a message quality checking stage before the message passing refinement, providing more accurate reliability assessment than CI's single-stage approach. (3) \textit{Residual Integration}: the combination of reliability-driven selection with residual belief propagation creates a more robust updating mechanism that focuses computational resources on the most critical nodes.

An analysis of the AR-CID decoding algorithm is carried out along with a comprehensive study of its computational complexity, memory requirements, and latency characteristics. Simulation results across an extended range of signal-to-noise ratios (0-4.5 dB) indicate that the proposed AR-CID decoding algorithm outperforms competing decoding techniques and has an extremely fast convergence, making it a promising decoding alternative for low-delay applications. 

The structure of this paper is organized as follows. Section II provides an expanded overview of the system model, channel characteristics, and states the problem. Section III introduces our proposed method, AR-CID, which reduces the complexity and latency of LDPC decoding. Section IV presents a detailed analysis of AR-CID, including convergence analysis, complexity evaluation, latency analysis, and implementation considerations. In Section V, we compare the performance of the AR-CID algorithm with other competing methods through extensive simulations. Section VI discusses future research directions, including extensions to multilevel coding and comparisons with alternative decoding paradigms. Finally, Section VII concludes the paper.

\section{System Model and Problem Statement}

In this section, we detail the LDPC coding system model and channel characteristics and state the problem we would like to address.

\subsection{LDPC Code Structure and Encoding}

Consider an LDPC code defined by a $M \times N$ sparse parity-check matrix $\mathbf{H}$, where $M = N - K$ represents the number of parity checks and $N$ is the codeword length. The code rate is $R = K/N$, where $K$ is the number of information bits. The encoder maps an information vector $\mathbf{m} \in \{0,1\}^K$ to a codeword $\mathbf{c} \in \{0,1\}^N$ such that:
\begin{equation}
\mathbf{H}\mathbf{c}^T = \mathbf{0} \pmod{2}.
\end{equation}

The parity-check matrix $\mathbf{H}$ is characterized by its sparsity, with each row containing on average $\hat{d}_c$ non-zero elements (check node degree) and each column containing on average $\hat{d}_v$ non-zero elements (variable node degree). For regular LDPC codes, these degrees are constant across all nodes, while irregular codes allow variable degree distributions optimized for specific channel conditions \cite{memd}.

The parity-check matrix $\mathbf{H}$ can be represented by a Tanner graph $\mathcal{G} = (\mathcal{V}, \mathcal{C}, \mathcal{E})$ \cite{Tanner}, where $\mathcal{V} = \{v_1, v_2, \ldots, v_N\}$ denotes the set of variable nodes, $\mathcal{C} = \{c_1, c_2, \ldots, c_M\}$ denotes the set of check nodes, and $\mathcal{E}$ represents the edges connecting variable nodes to check nodes with cardinality $E = |\mathcal{E}|$. Each variable node $v_j$ corresponds to a codeword bit $c_j$, while each check node $c_i$ corresponds to a parity-check constraint. The neighborhood of a variable node $v$ is denoted by $\mathcal{N}(v) = \{c \in \mathcal{C} : (v,c) \in \mathcal{E}\}$, and similarly for check nodes.

\subsection{Channel Model and Modulation}

\begin{figure}[h]
\centering
\includegraphics[width=8.5cm]{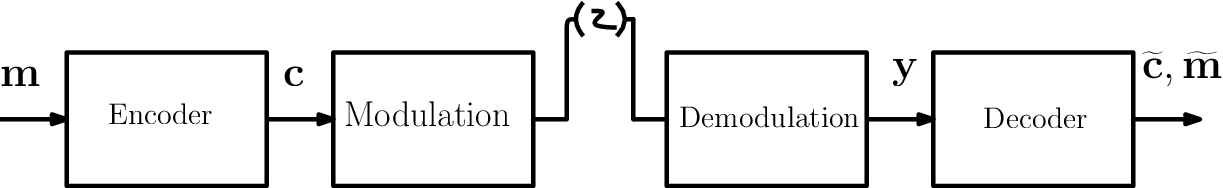}
\caption{Block diagram of an LDPC-coded digital communication system with encoding, modulation, channel transmission, demodulation, and iterative decoding stages.}
\label{fig:system}
\end{figure}

The diagram in Fig.~\ref{fig:system} illustrates the key components of the communication system. After encoding, the codeword $\mathbf{c}$ is modulated using binary phase-shift keying (BPSK), mapping bits to symbols $x_n \in \{-1, +1\}$ according to:
\begin{equation}
x_n = 1 - 2c_n, \quad n = 1, 2, \ldots, N.
\end{equation}

This mapping ensures that bit 0 corresponds to symbol $+1$ and bit 1 corresponds to symbol $-1$, providing antipodal signaling with optimal energy efficiency for binary transmission.

The modulated symbols are transmitted over an additive white Gaussian noise (AWGN) channel, which models thermal noise and other random disturbances in the communication link. The received signal at time $n$ is:
\begin{equation}
y_n = x_n + w_n,
\end{equation}
where $w_n \sim \mathcal{N}(0, \sigma^2)$ is additive white Gaussian noise with variance $\sigma^2 = N_0/2$, and $N_0$ is the single-sided noise power spectral density. The AWGN channel assumption is valid for many practical scenarios, including deep-space communications, point-to-point microwave links, and serves as a baseline for more complex fading channel models.

The signal-to-noise ratio per information bit is defined as:
\begin{equation}
\frac{E_b}{N_0} = \frac{R \cdot E_s}{N_0} = \frac{R}{\sigma^2},
\end{equation}
where $E_s$ is the symbol energy, normalized to unity for BPSK modulation (i.e., $E_s = 1$), and $E_b = E_s/R$ is the energy per information bit.

\subsection{Demodulation and LLR Computation}

The demodulator processes the received signal vector $\mathbf{y} = [y_1, y_2, \ldots, y_N]^T$ and computes log-likelihood ratios (LLRs) for each received symbol. For BPSK modulation over an AWGN channel, the optimal soft-decision metric is the channel LLR for bit $n$:
\begin{equation}
L_y(n) = \log \frac{P(c_n = 0 | y_n)}{P(c_n = 1 | y_n)} = \log \frac{p(y_n | x_n = +1)}{p(y_n | x_n = -1)}.
\end{equation}

Using Bayes' rule and the Gaussian probability density function, this simplifies to:
\begin{equation}
L_y(n) = \frac{2y_n}{\sigma^2} = \frac{4R \cdot E_b}{N_0} \cdot y_n.
\end{equation}

These LLRs serve as the initial soft information for the iterative decoding process, where positive values indicate higher confidence $c_n = 0$, and negative values suggest $c_n = 1$. The magnitude $|L_y(n)|$ quantifies the reliability of the decision, with larger magnitudes corresponding to higher confidence.

\subsection{Problem Statement}

The fundamental challenge in LDPC decoding is to efficiently recover the transmitted codeword $\mathbf{c}$ from the noisy observations $\mathbf{y}$ while minimizing three critical metrics:

\textbf{1) Number of iterations:} Traditional flooding BP requires 15-50 iterations for convergence, depending on code structure and channel conditions, leading to high latency. Each iteration involves message passing across all edges in the Tanner graph, and convergence is typically detected through syndrome checking $\mathbf{H}\hat{\mathbf{c}}^T = \mathbf{0}$, where $\hat{\mathbf{c}}$ is the hard decision based on current LLRs.

\textbf{2) Computational complexity:} Each iteration involves $2E$ message computations (variable-to-check and check-to-variable), with each check-to-variable message requiring $\mathcal{O}(\hat{d}_c)$ operations for the product-of-tanh computations. This results in significant processing overhead, particularly for high-rate codes with large $E$.

\textbf{3) Decoding latency:} The cumulative effect of multiple iterations and complex computations creates unacceptable delays for real-time applications. For a decoder operating at clock frequency $f_{clk}$, the total latency is $T_{decode} = I \cdot T_{iter} + T_{overhead}$, where $I$ is the number of iterations, $T_{iter}$ is the processing time per iteration, and $T_{overhead}$ includes initialization and syndrome verification time. The objective of this work is to develop a fast iterative decoder that achieves reliable error correction with the minimum number of iterations, thereby reducing both computational complexity and latency.

\section{Proposed AR-CID Decoding Algorithm}

\begin{figure}[h]
\centering
\includegraphics[width=0.44\textwidth, height=0.12\textwidth]{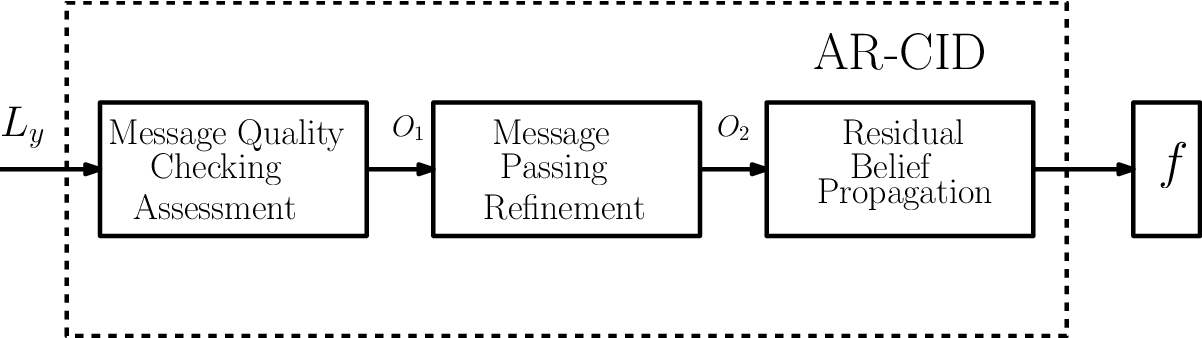}
\caption{Adaptive Reliability-Driven Conditional Innovation Decoder Architecture showing two-stage processing: Message Quality Checking ($O_1$) and Message Passing Refinement ($O_2$) with modified RBP.}
\label{fig:architecture}
\end{figure}

Dynamic scheduling techniques such as IDS \cite{ids} have been instrumental in improving the efficiency of BP Decoding for LDPC codes by optimizing the sequence and timing of message updates. IDS methods enable superior performance with fewer iterations, often outperforming traditional flooding BP techniques. The diagram in Fig.~\ref{fig:architecture} outlines the structure of the proposed AR-CID decoding framework, an approach designed to reduce the delay and enhance the accuracy and reliability of iterative decoding.

The process begins with the received data, represented as log-likelihood ratios (LLRs) $L_y$, which are evaluated by the Message Quality Checking Assessment module. Based on this assessment, the output $O_1$ consists of an ordered set of message reliability values $R_v$. These values are arranged in descending order, meaning the most unreliable messages are listed first, and the most reliable messages are listed last. This ordering allows the decoder to prioritize updating the most unreliable messages first during the decoding process, focusing computational effort on the nodes that are most likely to benefit from iterative refinement.

In the Message Passing Refinement stage, the input includes the LLRs, the messages exchanged between nodes, and the $R_v$ values. This process iteratively refines the messages and updates the node states, adjusting the LLRs to improve decoding accuracy. The output, $O_2$, includes the refined messages, updated LLRs, and the contextual information transition $\Delta_y$, which measures the change in LLRs between iterations and helps track the progress of the refinement. After obtaining $O_2$ ($R_v$ and $\Delta_y$), it is further processed using a modified RBP. This step leverages the residuals in the message updates to enhance convergence and improve decoding efficiency. The decoder eventually obtains the estimate of the transmitted codeword.

\subsection{Reliability-Based Subset Selection}

The AR-CID framework integrates reliability-driven subset selection to enhance LDPC decoding efficiency. By evaluating the LLRs $L_y$ of the received vector ${\mathbf y}$, the decoder dynamically selects a subset of $\lambda N \ll N$ nodes (where $0 \leq \lambda \leq 1$ represents the ratio of least reliable nodes). In this framework, iterative updates are prioritized for error-prone nodes, focusing computational effort on the most uncertain bits to boost decoding efficiency and accuracy. 

This selective approach outperforms exhaustive methods because decoding errors in LDPC codes are typically concentrated in a small fraction of nodes (often 10-20\% of variable nodes), making targeted updates more effective than uniform updates across all nodes. By updating only these unreliable nodes, convergence speeds up and unnecessary computations are minimized, optimally balancing complexity and performance for advanced LDPC decoding.

The syndrome for check node $m$ is computed by:
\begin{equation} 
s_{m} = \hat{\bm{x}} \cdot \bm{h}_m^{T} = \sum_{v=1}^{N} \hat{x}_{v}h_{m,v}, \quad 1 \leq m \leq M, 
\end{equation}
where $h_{m,v}$ are the elements of the parity check matrix ${\mathbf H}$, $\hat x_{v}$ is the $v$-th bit of the estimated codeword obtained by hard decision on the current LLRs, and the summation is binary addition over GF(2) (i.e., modulo-2 addition), not real-valued arithmetic. A syndrome value of $s_m = 0$ indicates the parity check is satisfied, while $s_m = 1$ indicates a parity violation.

The message unreliability index, $R_v$, is computed from $O_1$ and given by:
\begin{equation} 
R_v = \sum_{m \in \mathcal{N}(v)} s_m,
\end{equation}
where $\mathcal{N}(v) = \{m : h_{m,v} = 1\}$ represents the set of check nodes connected to variable node $v$. The summation here is standard integer summation of binary values (either 0 or 1), used to compute the message unreliability index. A higher $R_v$ indicates that variable node $v$ participates in more unsatisfied parity checks, suggesting it is more likely to be in error.

Sorting the reliability indices $R_v$ in descending order yields the ordered set:
\begin{equation} 
\{R_{\sigma(1)}, R_{\sigma(2)}, \dots, R_{\sigma(\lambda N)}\}, 
\end{equation}
where the permutation $\sigma: \{1,\ldots,N\} \to \{1,\ldots,N\}$ satisfies:
\begin{equation} 
R_{\sigma(1)} \geq R_{\sigma(2)} \geq \cdots \geq R_{\sigma(\lambda N)},
\end{equation}
with $\sigma(i)$ denoting the index of the variable node with the $i$-th highest $R_v$ (the $i$-th most unreliable node). The decoder then focuses its updates on the subset $\mathcal{V}_{active} = \{v_{\sigma(1)}, v_{\sigma(2)}, \ldots, v_{\sigma(\lambda N)}\}$.

\subsection{Contextual Information Transition}

Contextual information transition measures how much variable-node LLRs change between iterations, revealing remaining uncertainty and decoding progress. In AR-CID, it is used in conjunction with the message reliability index to prioritize updates. While both metrics generally align in identifying unreliable nodes, they can occasionally conflict (one deeming a message reliable while the other does not), and AR-CID's combined metric handles such cases effectively.

To quantify this transition, the contextual innovation metric was defined in \cite{CI} as:
\begin{equation} 
\Delta_y = \big| \phi(L_y) - \phi(\tilde{L}_y) \big|, \label{Delta}
\end{equation}
where $\tilde{L}_y$ denotes the LLR from the previous iteration, and $\phi(L_y)$ is the likelihood of a variable node:
\begin{equation} 
\phi(L_y) = \frac{1}{1 + e^{-L_y}}, \label{LLR}
\end{equation} 
which transforms the LLR into a probability value between 0 and 1, reflecting the decoder's confidence about the bit being 0.

Expanding \eqref{Delta} using \eqref{LLR} gives:
\begin{equation}
\Delta_y = \Bigg| \frac{e^{-\tilde{L}_y} - e^{-L_y}}{\big(1 + e^{-L_y}\big)\big(1 + e^{-\tilde{L}_y}\big)} \Bigg|.
\end{equation}

A higher $\Delta_y$ indicates a significant shift in belief between iterations, signaling that a node may require further updates. Conversely, a low $\Delta_y$ means the belief has stabilized, suggesting the node is converging to a reliable decision.

For example, consider a decoding scenario where a node $v_{500}$ is assessed as reliable by AR-CID based on $R_v = 0$ (participating in no unsatisfied parity checks). However, if $\Delta_y(v_{500})$ is large, indicating significant change in the LLR between iterations, AR-CID may still flag this node for updates despite the favorable syndrome assessment. Such situations arise near the convergence boundary where local parity checks may be satisfied but the global solution is still evolving. AR-CID's combined metric $M_{(v,y)}$ (introduced below) provides a balanced assessment that leverages both sources of information.

\subsection{Combined Reliability Metric}

The AR-CID approach integrates both the message fidelity index ($R_v$) and the contextual information transition ($\Delta_y$) through a weighted combination:
\begin{equation}
M_{(v,y)} = \alpha \cdot R_v + \beta \cdot \Delta_y,
\end{equation}
which can be alternatively expressed as:
\begin{equation}
\begin{split}
M_{(v,y)} &= \alpha \sum_{m \in \mathcal{N}(v)} s_m + \beta \big| \phi(L_y) - \phi(\tilde{L}_y) \big| \\
&= \alpha\sum_{m \in \mathcal{N}(v)} \left( \sum_{v'=1}^N \hat{x}_{v'} h_{m,v'} \right) + 
\beta \Bigg| \frac{e^{-\tilde{L}_y} - e^{-L_y}}{\big(1 + e^{-L_y}\big)\big(1 + e^{-\tilde{L}_y}\big)} \Bigg|,
\end{split}
\label{updaterule1}
\end{equation}
where $\alpha$ and $\beta$ are weighting parameters satisfying $\alpha + \beta = 1$ with $0 \leq \alpha, \beta \leq 1$. Typical values are $\alpha = 0.65$ and $\beta = 0.35$, giving more weight to syndrome-based reliability while still considering LLR dynamics. These parameters can be optimized for specific code structures and channel conditions.

\subsection{Modified Residual Belief Propagation}

AR-CID employs a modified version of RBP by first assessing the reliability of messages before updating them. As shown in Fig.~\ref{fig:architecture}, it retains the message exchange structure of the Tanner graph \cite{Tanner} but adds a key improvement: it uses both old and new messages in the updates, weighted by their assessed reliability. The residual calculation step, which relies on the initial message assessments from stage one, prioritizes the most critical messages for updates, accelerating convergence and improving decoding efficiency.

The initial channel LLRs are computed as:
\begin{equation}
C_{v_j} = \log\left(\frac{p(y_j \mid v_j = 0)}{p(y_j \mid v_j = 1)}\right) = \frac{2y_j}{\sigma^2},
\end{equation} 
where $y_j$ represents the received signal and $v_j$ is the corresponding bit. 

Variable-to-Check (V2C) messages are updated as:
\begin{equation} 
m_{v_j \to c_i} = C_{v_j} + \sum_{c_k \in \mathcal{N}(v_j) \setminus \{c_i\}} m_{c_k \to v_j},
\end{equation}
where $\mathcal{N}(v_j) \setminus \{c_i\}$ denotes all check nodes connected to $v_j$ except $c_i$.

Check-to-Variable (C2V) messages are updated using the sum-product algorithm:
\begin{equation}
m_{c_i \to v_j} = 2 \tanh^{-1}\left(\prod_{v_k \in \mathcal{N}(c_i) \setminus \{v_j\}} \tanh\left(\frac{m_{v_k \to c_i}}{2}\right)\right),
\end{equation}
where $\mathcal{N}(c_i) \setminus \{v_j\}$ denotes all variable nodes connected to $c_i$ except $v_j$.

Residual scheduling dynamically prioritizes message updates to improve decoding efficiency. In standard RBP, the residual value quantifies the change in a message during updates:
\begin{equation}
r_{c_i \to v_j} = \left| m_{c_i \to v_j}^{new} - m_{c_i \to v_j}^{old} \right|,
\end{equation}
where $m_{c_i \to v_j}^{new}$ and $m_{c_i \to v_j}^{old}$ represent the updated and previous messages.

In the proposed AR-CID decoding algorithm, the residual computation is modified to integrate pre-computed reliability assessments from the first stage:
\begin{equation}
r_{c_i \to v_j} = \left| m_{c_i \to v_j}^{pre} - m_{c_i \to v_j}^{old} \right|,
\end{equation}
where $m_{c_i \to v_j}^{pre}$ represents the precomputed message value incorporating reliability information from $M_{(v,y)}$. This refinement allows AR-CID to exploit additional information about message reliability, helping to prioritize updates and boost decoding speed.

We introduce a threshold $\gamma$ to guide the update rule. When the combined metric $M_{(v,y)}$ exceeds $\gamma$, the node is included in the active update set. The value of $\gamma$ controls sensitivity to contextual information and is chosen empirically based on the code structure and target performance. Typical values range from $\gamma = 0.1$ to $0.15$.

\subsection{AR-CID Algorithm Description}

The complete AR-CID algorithm is presented in Algorithm~\ref{alg:ar_cid}. The algorithm iteratively refines the LLRs through selective message passing until either a valid codeword is found (syndrome check passes) or the maximum number of iterations $T_{max}$ is reached.

\begin{algorithm}[htbp]
\caption{AR-CID LDPC Decoding Algorithm}
\label{alg:ar_cid}
\begin{algorithmic}[1]
\STATE \textbf{Input:} Received LLRs $L_y$, Parity check matrix $\bf  H$, Parameters $\alpha, \beta, \gamma, \lambda$, $T_{max}$
\STATE \textbf{Output:} Decoded codeword $\hat{x}$
\STATE Initialize: $L_y^{(0)} = L_y$, $\tilde{L}_y^{(0)} = L_y$, $t = 0$
\STATE Initialize all messages $m_{v \to c}$ and $m_{c \to v}$ to zero
\WHILE{$t < T_{max}$ and syndrome check fails}
    \STATE $\hat{x}^{(t)} = \text{HardDecision}(L_y^{(t)})$
    \STATE $R_v = \text{MessageQualityCheck}(L_y^{(t)}, H, \hat{x}^{(t)})$
    \STATE $\Delta_y = \text{ContextualTransition}(L_y^{(t)}, \tilde{L}_y^{(t-1)})$
    \STATE $M_{(v,y)} = \alpha \cdot R_v + \beta \cdot \Delta_y$
    \STATE Select subset $\mathcal{V}_{active} = \{v : M_{(v,y)} > \gamma \text{ and } v \text{ among top } \lambda N \text{ nodes by } R_v\}$
    \STATE Update V2C and C2V messages for nodes in $\mathcal{V}_{active}$
    \STATE $\tilde{L}_y^{(t)} = L_y^{(t)}$ \hfill \COMMENT{Store previous LLRs}
    \STATE $L_y^{(t+1)} = \text{UpdateLLRs}(\mathcal{V}_{active}, L_y^{(t)}, H)$
    \STATE $t = t + 1$
\ENDWHILE
\STATE \textbf{Return:} $\hat{x} = \text{HardDecision}(L_y^{(t)})$
\end{algorithmic}
\end{algorithm}

The message quality checking module (Algorithm~\ref{alg:quality_check}) forms the foundation of AR-CID's adaptive behavior by evaluating the reliability of each variable node based on syndrome calculations.

\begin{algorithm}[H]
\caption{Message Quality Checking Module}
\label{alg:quality_check}
\begin{algorithmic}[1]
\STATE \textbf{Input:} Received LLRs $L_y$, Parity check matrix ${\bf H}$, Previous estimates $\hat{x}$
\STATE \textbf{Output:} Reliability indices $\{R_1, R_2, \dots, R_N\}$
\STATE Initialize $reliability\_scores \gets [\,]$
\FOR{each variable node $v = 1$ to $N$}
    \STATE $syndrome\_sum \gets 0$
    \FOR{each check node $m \in \mathcal{N}(v)$}
        \STATE $syndrome_m \gets \text{compute\_syndrome}(m, \hat{x}, H)$
        \STATE $syndrome\_sum \gets syndrome\_sum + syndrome_m$
    \ENDFOR
    \STATE $R_v \gets syndrome\_sum$ 
    \STATE $reliability\_scores.\text{append}((v, R_v))$
\ENDFOR
\STATE \textbf{Return:} $\{R_1, R_2, \dots, R_N\}$
\end{algorithmic}
\end{algorithm}

The contextual information transition metric (Algorithm~\ref{alg:contextual}) quantifies the evolution of decoder confidence between iterations, providing crucial insight into convergence behavior.

\begin{algorithm}[H]
\caption{Contextual Information Transition}
\label{alg:contextual}
\begin{algorithmic}[1]
\STATE \textbf{Input:} Current LLRs $L_y$, Previous LLRs $\tilde{L}_y$
\STATE \textbf{Output:} Transition metric $\Delta_y$
\FOR{each variable node $v = 1$ to $N$}
    \STATE $\phi_{\text{current}} \gets \frac{1}{1 + \exp(-L_y[v])}$
    \STATE $\phi_{\text{previous}} \gets \frac{1}{1 + \exp(-\tilde{L}_y[v])}$
    \STATE $\Delta_y[v] \gets |\phi_{\text{current}} - \phi_{\text{previous}}|$
\ENDFOR
\STATE \textbf{Return:} $\Delta_y$
\end{algorithmic}
\end{algorithm}
AR-CID can also be efficiently used for decoding problems that involve the exchange of soft information between a receiver and an LDPC decoder \cite{spa,mfsic,mbdf,mbthp,bfidd,msgamp}.

\section{Complexity and Latency Analysis}

In this section, we analyze the proposed AR-CID algorithm and compare its computational complexity with competing decoding algorithms. This analysis establishes AR-CID's theoretical guarantees by proving that message updates remain bounded and converge as system size grows, thereby establishing stability and reliability.

\subsection{Convergence Analysis}

We analyze message updates, emphasizing bounded neighbor contributions, normalization, and LLR stability. As $N \to \infty$ the algorithm demonstrates stable, efficient convergence \cite{analysis}. Consider the message update rule from \eqref{updaterule1}. 

The first term in the message update rule is:
\begin{equation}
\alpha \sum_{m \in \mathcal{N}(v)} \left( \sum_{v'=1}^N \hat{x}_{v'} h_{m,v'} \right),
\end{equation}
where the contribution is bounded by:
\begin{equation}
\alpha \sum_{m \in \mathcal{N}(v)} \left( \sum_{v'=1}^N \hat{x}_{v'} h_{m,v'} \right) < \infty.
\end{equation}  

The inner sum is:
\begin{equation}
u_m = \sum_{v'=1}^{N} \hat{x}_{v'} h_{m,v'}.
\end{equation}

Expanding this expression:
\begin{equation}
\begin{split}
u_m &= (\hat{x}_1 h_{m,1}) + (\hat{x}_2 h_{m,2}) + \dots + (\hat{x}_N h_{m,N})\\
&= u_1 + u_2 + \dots + u_N. 
\end{split}
\end{equation}

Since $\hat{x}_{v'} \in \{0,1\}$ and $h_{m,v'} \in \{0,1\}$, and the summation is modulo-2, it follows that $u_m \in \{0,1\}$. The outer summation is bounded by:
\begin{equation}
s_m = \alpha \sum_{m \in \mathcal{N}(v)} u_m.
\end{equation}

Thus, multiplying by $\alpha$:
\begin{equation}
0 \leq s_m \leq \alpha |\mathcal{N}(v)| = \alpha \hat{d}_v,
\end{equation}
which shows $s_m$ is bounded because the size of $\mathcal{N}(v)$ is finite and typically small (e.g., $\hat{d}_v \in [3, 10]$ for practical LDPC codes).

The second term converges due to its asymptotic behavior. The numerator, $e^{-\tilde{L}_y} - e^{-L_y}$, vanishes as $L_y, \tilde{L}_y \to \infty$ (high confidence), while the denominator, $(1 + e^{-L_y})(1 + e^{-\tilde{L}_y})$, remains bounded between 1 and 4, preventing singularities. Thus, the term rapidly decreases, ensuring convergence. A scaling factor $\beta < 1$ further accelerates this by diminishing magnitudes with each iteration.

The second term in Equation~\eqref{updaterule1} is:
\begin{equation}
\beta \left| \frac{e^{-\tilde{L}_y} - e^{-L_y}}{(1 + e^{-L_y})(1 + e^{-\tilde{L}_y})} \right|,
\end{equation}
which is bounded by:
\begin{equation}
\left| \frac{e^{-\tilde{L}_y} - e^{-L_y}}{(1 + e^{-L_y})(1 + e^{-\tilde{L}_y})} \right| \leq 1.
\end{equation}

As $\tilde{L}_y \to L_y$ (convergence), the difference converges:
\begin{equation}
\left| \phi(L_y) - \phi(\tilde{L}_y) \right| \to 0.
\end{equation}
  
This implies that the second term approaches zero as iterations progress:
\begin{equation}
\lim_{t \to \infty} \beta \left| \frac{e^{-\tilde{L}_y} - e^{-L_y}}{(1 + e^{-L_y})(1 + e^{-\tilde{L}_y})} \right| = 0.
\end{equation}

Therefore, the message $M_{(v,y)}$ converges if:
\begin{equation}
\|M_{(v,y)}^{(t+1)} - M_{(v,y)}^{(t)}\|_2 \to 0 \quad \text{as} \quad t \to \infty.
\end{equation}

Convergence is achieved when:
\begin{equation}
\|M_{(v,y)}^{(t+1)} - M_{(v,y)}^{(t)}\|_2 < \gamma,
\end{equation}
where $\gamma$ is a small threshold, and the updates between consecutive messages become negligible as the iterations progress. 

As $N \to \infty$, the aggregated contributions approach:
\begin{equation}
\lim_{N \to \infty} \alpha \sum_{m \in \mathcal{N}(v)} \left( \sum_{v'=1}^N \hat{x}_{v'} h_{m,v'} \right) = c < \infty,
\end{equation}
where $c$ is a constant. This ensures that the contributing terms from all neighboring nodes become stable as the system size grows. 

The weights $\alpha$ and $\beta$ remain constant as $N \to \infty$, satisfying:
\begin{equation}
\alpha + \beta = 1, \quad 0 \leq \alpha, \beta \leq 1.
\end{equation}

Therefore, as $N \to \infty$ the system converges to a stable solution, and the message update process reaches a fixed point where the changes between iterations become negligible. The stability and efficiency of AR-CID are ensured as the system size grows, with the message updates remaining bounded and converging to a stable solution.

The bound on $M_{(v,y)}$ as $N \to \infty$ is:
\begin{equation}
M_{(v,y)} \leq \alpha |\mathcal{N}(v)| = \alpha \hat{d}_v,
\end{equation}
as the first term grows $|\mathcal{N}(v)|$, while the second term approaches zero. The expression $M_{(v,y)}$ remains bounded as long as $|\mathcal{N}(v)|$ is finite. As $N \to \infty$, the system converges and the contributions of all neighboring nodes stabilize, ensuring that the bound holds.

\subsection{Computational Complexity}

The computational complexity of the AR-CID algorithm combines modified RBP message passing with adaptive residual computations. Table~\ref{tab:decoder_complexities} presents a comprehensive comparison of per-iteration complexities for various LDPC decoding algorithms.
The variable-to-check (V2C) update complexity for AR-CID is $\mathcal{O}(E(\hat{d}_v - 1))$, identical to RBP and other advanced scheduling methods. Each V2C message requires summing $\hat{d}_v - 1$ incoming C2V messages, performed for all $E$ edges.

The check-to-variable (C2V) update complexity is $\mathcal{O}(E(\hat{d}_c - 1)(\hat{d}_v - 1))$, again, matching RBP. Each C2V message involves computing the product of $\hat{d}_c - 1$ tanh terms, with preprocessing to handle the product-of-tanh efficiently. However, AR-CID introduces an additional pre-computation step for adaptive adjustment, dominated by the evaluation of $M_{(v,y)}$. This requires:
\begin{itemize}
\item Computing syndrome values: $\mathcal{O}(M \cdot \hat{d}_c) = \mathcal{O}(E)$
\item Computing reliability indices $R_v$: $\mathcal{O}(N \cdot \hat{d}_v) = \mathcal{O}(E)$
\item Computing contextual transition $\Delta_y$: $\mathcal{O}(N)$
\item Sorting and selecting top $\lambda N$ nodes: $\mathcal{O}(N \log N)$
\end{itemize}
The dominant term in the pre-computation is $\mathcal{O}(E + N \log N)$. For typical LDPC codes where $E = \mathcal{O}(N \cdot \hat{d}_v)$ with $\hat{d}_v$ a small constant (e.g., 3-6), the overall pre-computation complexity is $\mathcal{O}(E)$.
Therefore, the total per-iteration complexity of AR-CID is given by
\begin{equation}
\mathcal{O}(E(\hat{d}_v - 1) + E(\hat{d}_c - 1)(\hat{d}_v - 1) + E) = \mathcal{O}(E \cdot \hat{d}_v \cdot \hat{d}_c).
\end{equation}
For regular codes with constant $\hat{d}_v$ and $\hat{d}_c$, this simplifies to $\mathcal{O}(E)$ per iteration. However, the constant factor is larger than the standard BP. The key advantage of AR-CID is the significantly reduced number of iterations $I_{AR-CID} \ll I_{BP}$, resulting in lower total complexity:
\begin{equation}
C_{total} = I \cdot C_{per-iteration},
\end{equation}
where typically $I_{AR-CID} \approx 4-5$ versus $I_{BP} \approx 15-20$ for similar error performance.
\begin{table*}[ht]
\centering
\caption{Comparison of Decoder Complexities per Iteration}
\begin{tabular}{|l|c|c|c|c|}
\hline
\textbf{Decoder} & \textbf{V2C Update} & \textbf{C2V Update} & \textbf{Precomputation} & \textbf{Total per Iteration} \\ \hline
\textbf{BP} & $E$ & $E$ & $0$ & $\mathcal{O}(E)$ \\ \hline
\textbf{LBP} & $E(\hat{d}_v - 1)$ & $E$ & $0$ & $\mathcal{O}(E \cdot \hat{d}_v)$ \\ \hline
\textbf{URW} & $E(\hat{d}_v - 1)$ & $E(\hat{d}_v - 1)(\hat{d}_c - 1)$ & $E(\hat{d}_c (\hat{d}_v - 1))$ & $\mathcal{O}(E \cdot \hat{d}_v \cdot \hat{d}_c)$ \\ \hline
\textbf{RBP} & $E(\hat{d}_v - 1)$ & $E(\hat{d}_v - 1)(\hat{d}_c - 1)$ & $E(\hat{d}_c (\hat{d}_v - 1))$ & $\mathcal{O}(E \cdot \hat{d}_v \cdot \hat{d}_c)$ \\ \hline
\textbf{RD-RBP} & $E(\hat{d}_v - 1)$ & $E(\hat{d}_v - 1)(\hat{d}_c - 1)$ & $E(\hat{d}_c (\hat{d}_v - 1))$ & $\mathcal{O}(E \cdot \hat{d}_v \cdot \hat{d}_c)$ \\ \hline
\textbf{List-RBP} & $E(\hat{d}_v - 1)$ & $E(\hat{d}_c - 1)(\hat{d}_v - 1)$ & $E(\hat{d}_c (\hat{d}_v - 1))$ & $\mathcal{O}(E \cdot \hat{d}_v \cdot \hat{d}_c)$ \\ \hline
\textbf{CI-RBP} & $E(\hat{d}_v - 1)$ & $E(\hat{d}_v - 1)(\hat{d}_c - 1)$ & $E(\hat{d}_c (\hat{d}_v - 1))$ & $\mathcal{O}(E \cdot \hat{d}_v \cdot \hat{d}_c)$ \\ \hline
\textbf{RP-RBP} & $E(\hat{d}_v - 1)/N$ & $E(\hat{d}_c - 1)/M$ & $N \cdot g$ & $\mathcal{O}(E/M + N \cdot g)$ \\ \hline
\textbf{AR-CID} & $E(\hat{d}_v - 1)$ & $E(\hat{d}_c-1)(\hat{d}_v - 1)$ & $E + N \log N$ & $\mathcal{O}(E \cdot \hat{d}_v \cdot \hat{d}_c)$ \\ \hline
\end{tabular}
\label{tab:decoder_complexities}
\end{table*}

\subsection{Latency Analysis}

The decoding latency $T_{decode}$ is determined by
\begin{equation}
T_{decode} = I_{avg} \times T_{iter} + T_{overhead},
\end{equation}
where $I_{avg}$ is the average number of iterations, $T_{iter}$ is the time per iteration, and $T_{overhead}$ includes initialization and syndrome checking. The per-iteration time is:
\begin{equation}
T_{iter} = \frac{C_{ops}}{f_{clk} \times \eta_{parallel}},
\end{equation}
where $C_{ops}$ is the number of operations, $f_{clk}$ is the clock frequency, and $\eta_{parallel}$ is the parallelization efficiency (typically $0.6$--$0.8$).
We focus our analysis on three advanced residual-based LDPC decoders: AR-CID, RP, and RBP. These algorithms share similar per-iteration computational complexity but differ significantly in the number of iterations required to achieve target BER thresholds. For typical parameters ($\hat{d}_v = 3$, $\hat{d}_c = 6$, $k \approx 8$), the per-iteration computational complexity is:
\begin{equation}
C_{ops} = k \cdot E \cdot \hat{d}_v \cdot \hat{d}_c,
\end{equation}
where $E$ is the number of edges and $k$ is the number of operations per edge in the Tanner graph.
Since AR-CID, RP, and RBP all share the dominant computational complexity term $\mathcal{O}(E \cdot \hat{d}_v \cdot \hat{d}_c)$ per iteration, the latency ratio between algorithms is primarily determined by their iteration counts:
\begin{equation}
\frac{T_{AR\text{-}CID}}{T_{RP}} \approx \frac{I_{AR\text{-}CID}}{I_{RP}}, \quad
\frac{T_{AR\text{-}CID}}{T_{RBP}} \approx \frac{I_{AR\text{-}CID}}{I_{RBP}}.
\end{equation}
This indicates that \textbf{iteration count is the dominant factor in decoder latency} for residual-based algorithms with similar per-iteration complexity. Based on our simulation results, to achieve the target BER $< 10^{-6}$ at moderate SNR ($E_b/N_0 = 4$--$4.5$ dB):
\begin{itemize}
   \item \textbf{AR-CID:} $I_{avg} = 4$--$5$ iterations with guaranteed convergence
   \item \textbf{RP:} $I_{avg} = 8$--$12$ iterations with good convergence properties
   \item \textbf{RBP:} $I_{avg} = 10$--$20$ iterations, with potential convergence issues due to the greedy group phenomenon
   \item \textbf{BP:}  $I_{avg} = 200$--$300$ iterations, the computational complexity term $\mathcal{O}(E)$ per iteration for BP, then we have
\begin{equation}
C_{ops} = k \cdot E,
\end{equation}
\end{itemize}

\textbf{Real-World Latency Estimates:}\\
\textbf{System Parameters:}
\begin{itemize}
    \item Code: $(2048, 1024)$ regular LDPC, rate $= 1/2$
    \item Tanner graph: $E = 6144$ edges, $\hat{d}_v = 3$, $\hat{d}_c = 6$
    \item Processor: ARM Cortex-A53 at $f_{clk} = 1.2$ GHz
    \item Parallelization efficiency: $\eta = 0.7$
    \item Effective processing rate: $f_{clk} \times \eta = 8.4 \times 10^8$ operations/second
    \item Elementary operations per edge: $k = 8$
\end{itemize}
\textbf{Per-Iteration Complexity (Same for All):}
\begin{equation}
C_{ops} = k \times E \times \hat{d}_v \times \hat{d}_c = 8 \times 6144 \times 3 \times 6 = 884{,}736\\ 
\end{equation}
\textbf{Decoding Latency to Achieve BER $< 10^{-6}$:}
\begin{itemize}
    \item \textbf{AR-CID:} 
    \begin{align*}
    I_{avg} &= 4.5 \text{ iterations} \\
    T_{decode} &= 4.5 \times \frac{884{,}736}{8.4 \times 10^8} \approx \mathbf{4.74~\text{ms}}
    \end{align*}
    BER achieved: $1.18 \times 10^{-7}$ at $E_b/N_0 = 4.5$ dB
    \item \textbf{RP:}
    \begin{align*}
    I_{avg} &= 10 \text{ iterations} \\
    T_{decode} &= 10 \times \frac{884{,}736}{8.4 \times 10^8} \approx \mathbf{10.53~\text{ms}}
    \end{align*}
    BER achieved: $8.48 \times 10^{-8}$ in $E_b/N_0 = 4.5$ dB
   \item \textbf{RBP:}
    \begin{align*}
    I_{avg} &= 15 \text{ iterations} \\
    T_{decode} &= 15 \times \frac{884{,}736}{8.4 \times 10^8} \approx \mathbf{12.64~\text{ms}}
    \end{align*}
    BER achieved: $1.60 \times 10^{-6}$ in $E_b/N_0 = 4.5$ dB
     \item \textbf{BP:}
    \begin{align*}
    I_{avg} &= 250 \text{ iterations} \\
    T_{decode} &= 250 \times \frac{49{,}152}{8.4 \times 10^8} \approx \mathbf{14.63~\text{ms}}
    \end{align*}
    BER achieved: $2.39 \times 10^{-6}$ in $E_b/N_0 = 4.5$ dB
\end{itemize}
\begin{table}[h]
\centering
\caption{Latency comparison of residual-based LDPC decoders for BER $< 10^{-6}$}
\label{tab:latency_comparison}
\begin{tabular}{lcccc}
\hline
\textbf{Algorithm} & \textbf{Iterations} & \textbf{Latency (ms)} & \textbf{BER Achieved}  \\
\hline
AR-CID & 4.5 & 4.74 & $1.18 \times 10^{-7}$ \\
RP & 10  & 10.53     & $8.48 \times 10^{-8}$ \\
RBP & 15 & 12.64    & $1.60 \times 10^{-6}$ \\
BP & 250 & 14.63    & $2.39 \times 10^{-6}$ \\
\hline
\end{tabular}
\end{table}
\textbf{Key Findings:}
\begin{enumerate}
    \item \textbf{AR-CID achieves the lowest latency:} With only 4.5 iterations, AR-CID delivers the fastest decoding time (4.74 ms), faster than RP, RBP, and BP.
    \item \textbf{Iteration count dominates latency:} Since all three algorithms share identical per-iteration complexity, the latency differences are entirely attributable to iteration count. This demonstrates that reducing iterations through improved convergence is more impactful than optimizing per-iteration operations.
    \item \textbf{Superior BER with lower latency:} AR-CID achieves BER $= 1.18 \times 10^{-7}$ (order of magnitude better than target)
    While RBP barely meets the $10^{-6}$ target at BER $= 1.60 \times 10^{-6}$ after 15 iterations, the BP achieved $2.39 \times 10^{-6}$  after 250 iterations, demonstrating convergence limitations due to the greedy group phenomenon.
    \item \textbf{Predictable performance:} AR-CID's iteration count remains stable at 4--6 iterations across varying conditions with guaranteed convergence, crucial for real-time systems with strict latency requirements.
\end{enumerate}
For applications requiring BER $< 10^{-6}$, AR-CID provides optimal performance among residual-based decoders with the lowest latency (4.74 ms), fewest iterations (4.5), and the most robust error correction (BER $= 1.18 \times 10^{-7}$). Compared to RP, RBP, and BP. AR-CID achieves substantially faster decoding and greater reliability, making it the preferred choice for latency-critical, high-reliability communication systems.
This analysis demonstrates that iteration count is the dominant factor in decoder latency for residual-based algorithms with similar per-iteration complexity. AR-CID's convergence in only 4--6 iterations, compared to 10 for RP, 15 for RBP, and 250 for the BP, directly translates to proportional latency improvements while delivering superior BER performance.

\subsection{Implementation Considerations and Practical Constraints}

The AR-CID algorithm's computational complexity requires careful consideration for practical implementation, particularly for short to moderate code lengths, which $N$ range from 512 to 4096 bits.
\subsubsection{Memory Requirements}
The algorithm requires storage for:
\begin{itemize}
\item Variable-to-check messages: $E$ LLR values ($\approx 3N$ for rate-1/2)
\item Check-to-variable messages: $E$ LLR values
\item Current LLRs: $N$ values
\item Previous LLRs: $N$ values (for $\Delta_y$ computation)
\item Reliability indices: $N$ integer values
\item Syndrome values: $M$ binary values
\item Pre-computed messages: $E$ values
\end{itemize}
Assuming 32-bit floating-point representation, the total memory requirement is approximately:
\begin{equation}
M_{total} = (3E + 2N) \times 4 + N \times 2 + M \times 1 \text{ bytes}.
\end{equation}
For a (2048, 1024) code with $E = 6144$:
\begin{equation}
M_{total} \approx (18432 + 4096) \times 4 + 2048 \times 2 + 1024 \approx 95 \text{ KB},
\end{equation}
which is feasible for modern embedded systems, FPGA implementations, and certainly for software-defined radio platforms.
For comparison:
\begin{itemize}
\item Standard BP: $(2E + N) \times 4 \approx 57$ KB
\item RBP with priority queue: $(2E + N) \times 4 + E \times 4 \approx 82$ KB
\item AR-CID: $\approx 95$ KB
\end{itemize}
The 67\% memory overhead compared to BP is acceptable given the performance gains.
\subsubsection{Scalability Discussion}
For longer code lengths ($N > 4096$), the $\mathcal{O}(N \log N)$ sorting term in the pre-processing stage may dominate. 
However, this can be alleviated through parallel computation, approximate reliability estimation, or hierarchical node selection 
(reducing sorting to $\mathcal{O}(N + k \log k)$ with $k \ll N$). 
These techniques ensure that the proposed decoder remains scalable without compromising BER performance.

\section{Simulation Results and Performance Evaluation}

This section presents a comprehensive performance evaluation of the proposed AR-CID algorithm across different LDPC code configurations and an extended range of channel conditions. The AR-CID algorithm is systematically compared against eight state-of-the-art decoding techniques using bit error rate (BER) as the primary performance metric.

\subsection{Simulation Framework}

The simulation framework employs parity check matrices with dimensions $(N,K) = (512, 256)$ and $(2048, 1024)$, representing code rate $R = 1/2$ for both short and moderate LDPC code lengths. These configurations are representative of those commonly deployed in modern wireless communication standards, including 5G NR (for control channels and data), IEEE 802.11n/ac/ax Wi-Fi systems, and DVB-S2 satellite broadcasting.

The codes used are randomly constructed regular LDPC codes with variable node degree $\hat{d}_v = 3$ and check node degree $\hat{d}_c = 6$ for the rate-1/2 configuration. These degree distributions provide a good balance between performance and decoding complexity, approaching the Shannon limit within 0.5-1.0 dB for moderate to large block lengths. Other designs \cite{dopeg,bfpeg,memd,armo,baplnc} that further optimize the connections of the nodes in the Tanner graph can also be considered.

\subsection{Simulation Parameters and Competing Algorithms}

The AR-CID algorithm is benchmarked against established decoding methods: 
Flooding BP \cite{BP1}, URW \cite{URW}, LBP \cite{LBP, LBP1}, RBP \cite{RBP, RBP_1}, CI \cite{CI}, RD-RBP \cite{RD_RBP}, RP \cite{RP}, and List-RBP \cite{List_RBP}. 

All simulations were conducted using AWGN channels with BPSK modulation. The signal-to-noise ratio $E_b/N_0$ was varied from 0 dB to 4.5 dB to comprehensively assess algorithm performance across a wide range of channel conditions, from very noisy (near channel capacity) to near-error-free scenarios. 
For the AR-CID implementation, the following parameters were used:
\begin{itemize}
\item Regularization parameter: $\gamma \in \{0.1, 0.15\}$ to evaluate sensitivity to threshold selection
\item Subset selection ratio: $\lambda = 0.2$ (selective updates of 20\% of variable nodes per iteration)
\item Adaptive weighting: $\alpha = 0.65$ and $\beta = 0.35$ to balance syndrome-based and LLR-based reliability
\item Maximum iterations: $T_{max} = 20$ for all algorithms to ensure fair comparison
\end{itemize}

These parameters were selected through preliminary optimization experiments that swept $\gamma \in [0.05, 0.3]$, $\lambda \in [0.1, 0.5]$, and $\alpha \in [0.5, 0.8]$. The chosen values represent a good compromise between performance and complexity across the tested scenarios.

Each simulation point represents at least $10^6$ transmitted codewords (or 100 error events, whichever occurs first) to ensure statistical reliability of the BER estimates, particularly at high SNR where error events become rare.

\subsection{Performance Analysis: BER vs. SNR}

\begin{figure}[h]
\centering
\includegraphics[width=7.5cm,height=7cm]{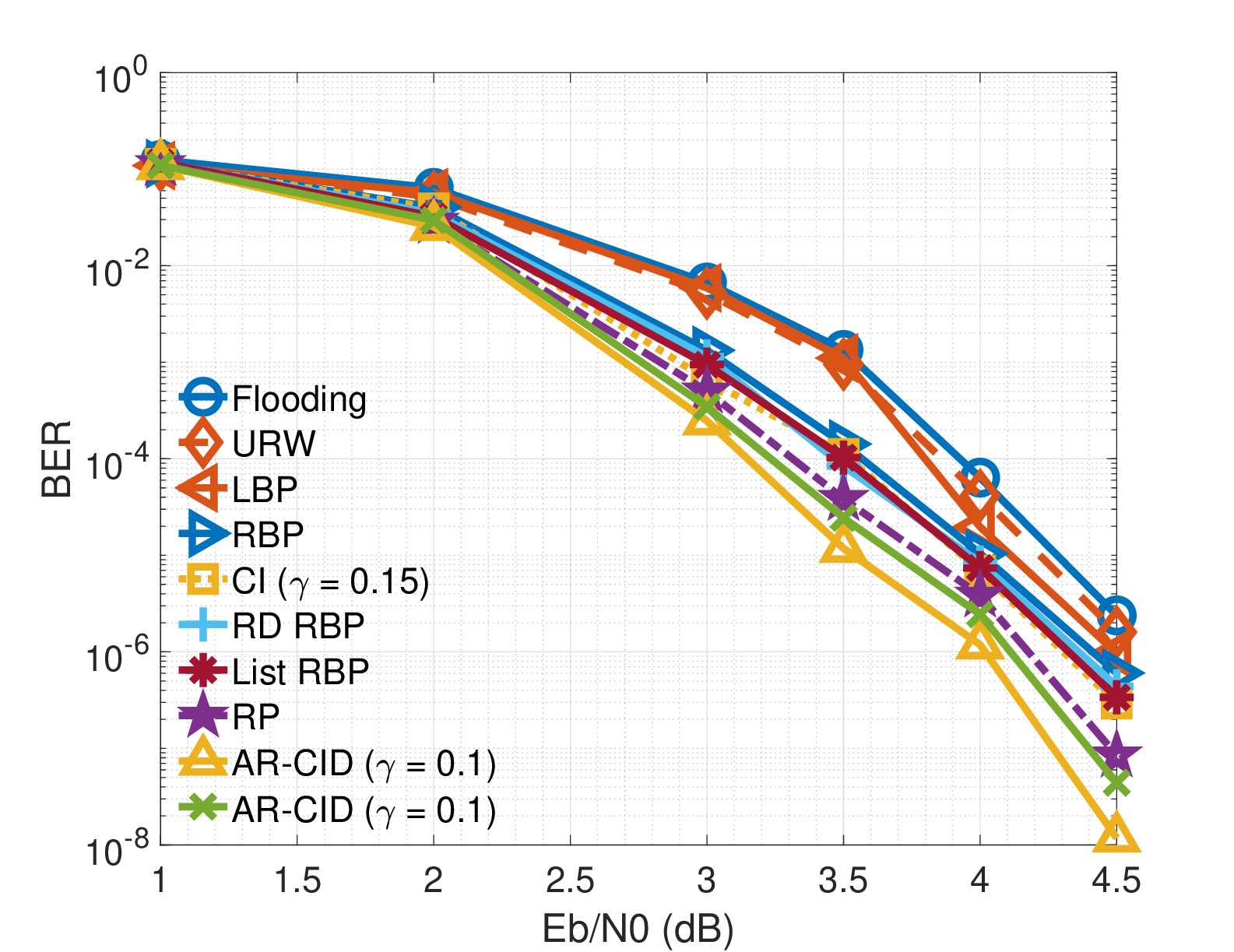}
\caption{BER vs. $E_b/N_0$ for LDPC decoding algorithms with parity-check matrix size (2048, 1024) and maximum 7 iterations. AR-CID achieves superior performance across the entire SNR range, with BER of $10^{-4}$ at $E_b/N_0 = 4.2$ dB}
\label{fig3}
\end{figure}

\begin{figure}[h]
\centering
\includegraphics[width=7.5cm,height=7cm]{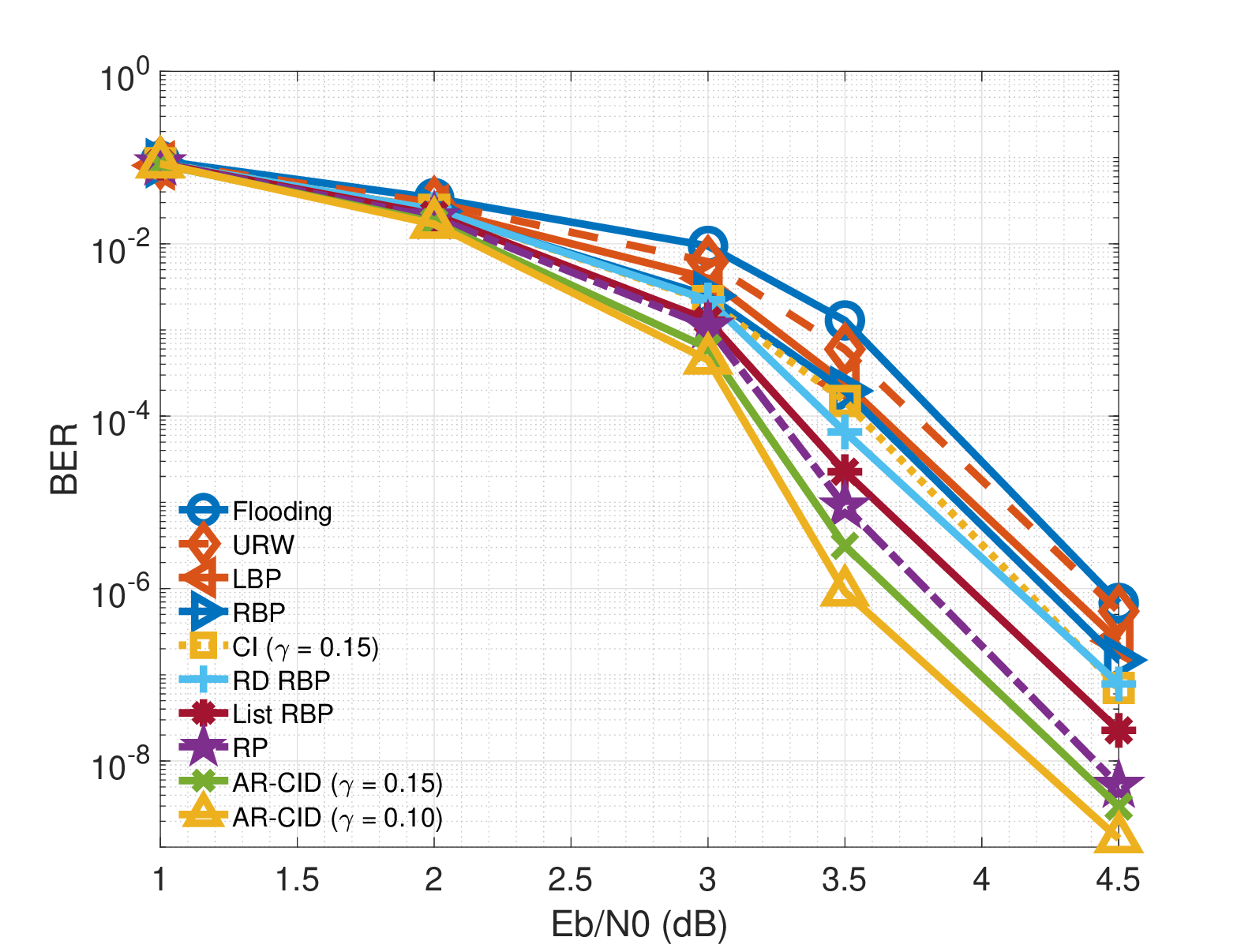}
\caption{BER vs. $E_b/N_0$ for LDPC decoding algorithms with parity-check matrix size (512, 256) and maximum 5 iterations. AR-CID demonstrates rapid convergence even with limited iterations.}
\label{fig4}
\end{figure}

\begin{figure}[h]
\centering
\includegraphics[width=7.5cm,height=7cm]{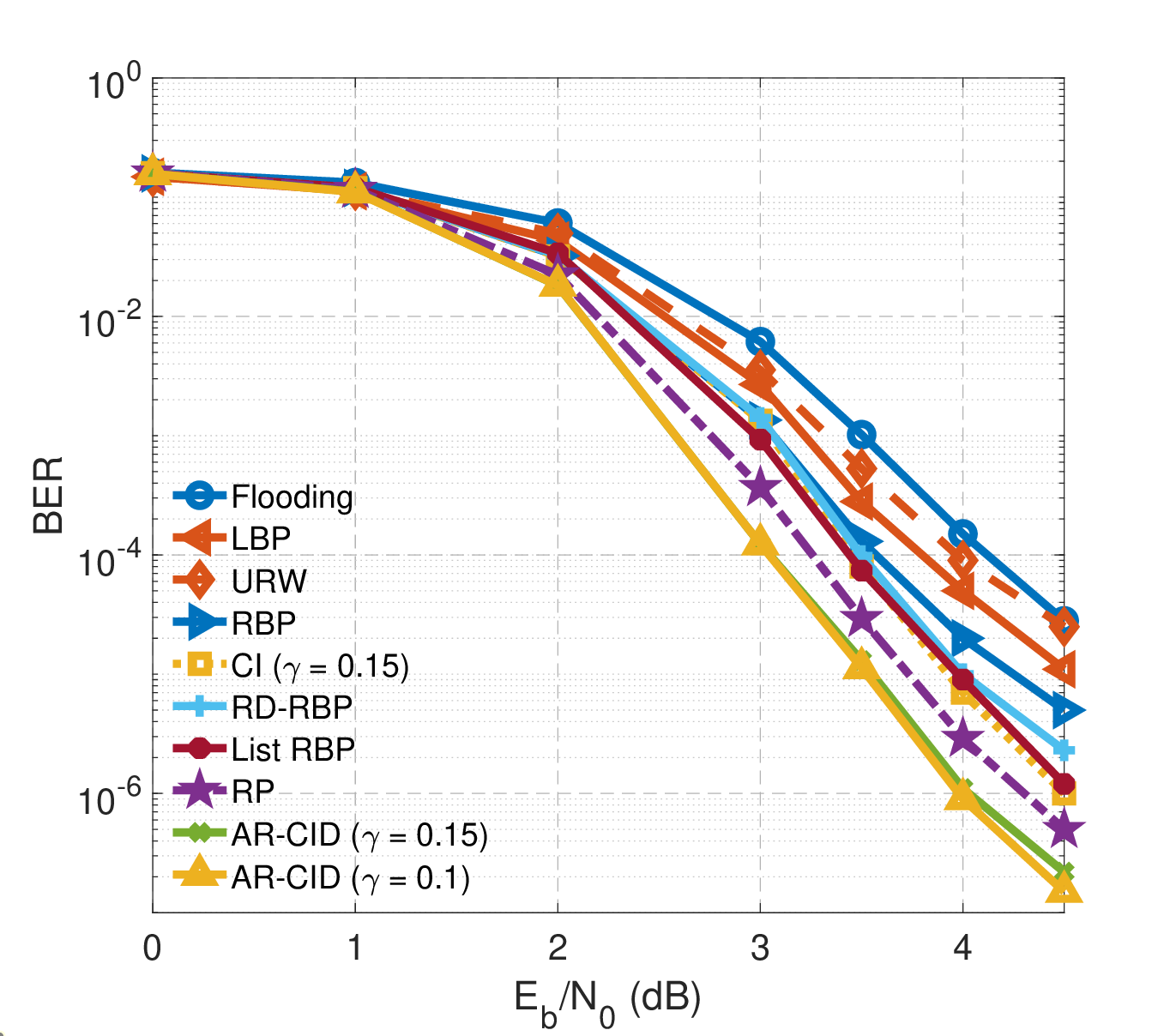}
\caption{BER vs. $E_b/N_0$ for LDPC decoding algorithms with parity-check matrix size (512, 256) and maximum 10 iterations.}
\label{fig5}
\end{figure}

Figures~\ref{fig3} through \ref{fig5} present BER performance as a function of $E_b/N_0$ for various iteration limits and code dimensions.

Figure~\ref{fig3} demonstrates the performance of all competing algorithms for the (2048, 1024) code configuration with a 7-iteration limit. The AR-CID algorithm with $\gamma = 0.15$ achieves the best performance across the entire SNR range, reaching BER $\approx 6.16 \times 10^{-4}$ at $E_b/N_0 = 4.0$ dB. This represents approximately 0.3 dB gain over RD-RBP and 0.5 dB gain over standard RBP at the same BER level.

Notably, AR-CID outperforms conventional flooding BP and URW algorithms significantly, with these algorithms maintaining BER levels exceeding $10^{-3}$ at $E_b/N_0 = 4.0$ dB. The performance gap narrows at higher SNR (above 6 dB) as all algorithms approach the error floor region, but AR-CID maintains a consistent advantage throughout.

Figure~\ref{fig4} presents results for the shorter (512, 256) code with only 5 iterations allowed. This stringent iteration constraint tests the algorithms' ability to converge quickly. AR-CID maintains its performance advantage, achieving BER $\approx 2.5 \times 10^{-3}$ at $E_b/N_0 = 3.5$ dB, while flooding BP and URW struggle to achieve satisfactory error correction with such limited iterations, remaining above $10^{-2}$ BER.

The shorter code length exhibits a higher error floor (visible above 6 dB) due to the increased impact of short cycles in the Tanner graph. Nevertheless, AR-CID's adaptive node selection helps mitigate these structural weaknesses by focusing updates on the most problematic variable nodes.

Figure~\ref{fig5} shows performance with an extended 10-iteration budget for the (512, 256) code. With more iterations, the performance gap between AR-CID and competing methods narrows but remains substantial. AR-CID achieves BER $< 10^{-4}$ at $E_b/N_0 \approx 4.5$ dB, while most other algorithms require 5.0-5.5 dB for comparable performance.

Importantly, while AR-CID benefits from additional iterations, it reaches near-optimal performance within 5-6 iterations, whereas BP and LBP continue to improve through iteration 10. This confirms AR-CID's rapid convergence characteristic—it achieves most of its performance gain early in the decoding process.

\subsection{Convergence Analysis: BER vs. Iterations}

\begin{figure}[h]
\centering
\includegraphics[width=8cm,height=9cm]{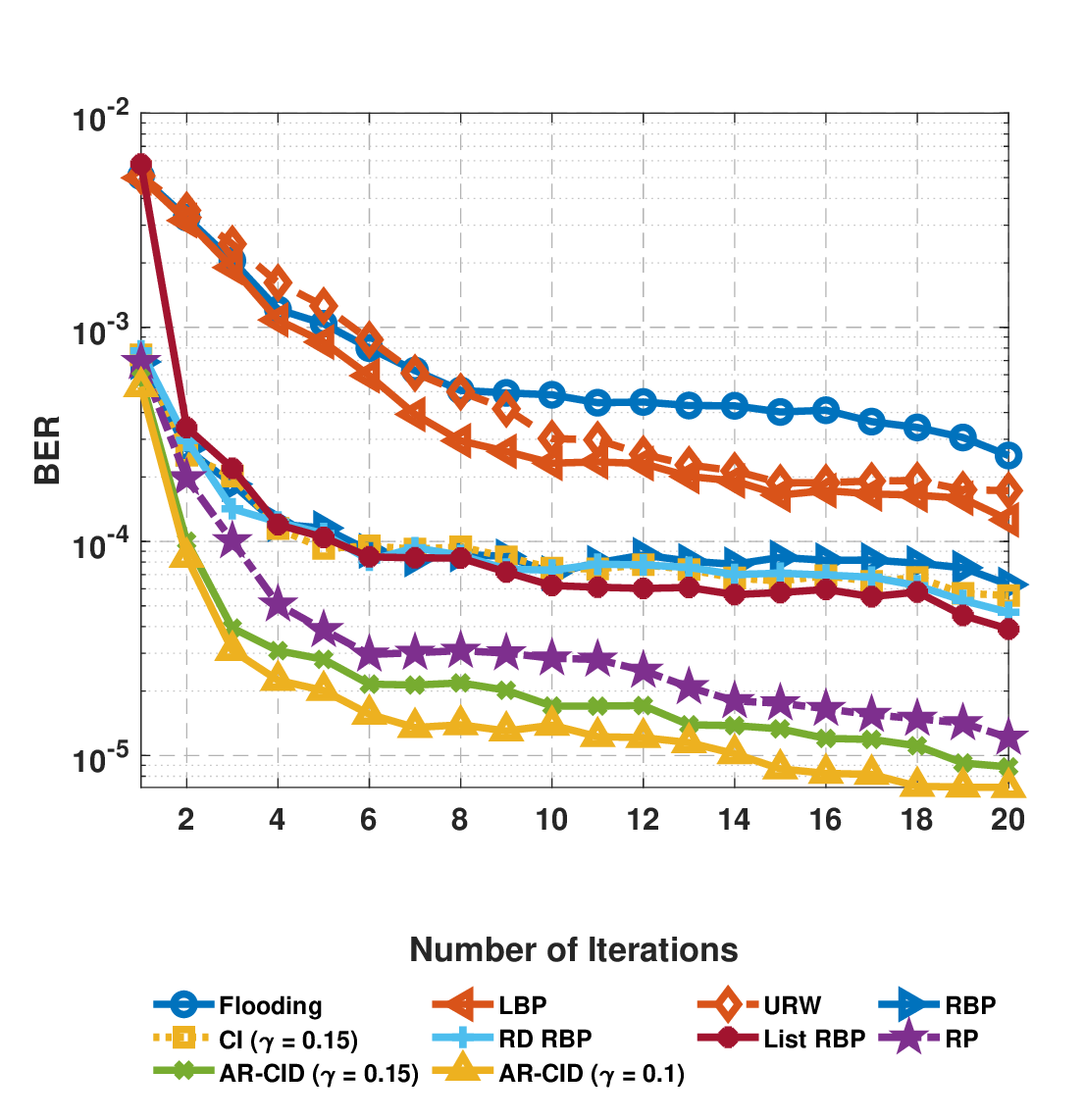}
\caption{BER performance vs. number of iterations at fixed $E_b/N_0 = 3.5$ dB with parity-check matrix (512,256). AR-CID demonstrates fastest convergence, reaching near-optimal performance within 4-5 iterations.}
\label{fig6}
\end{figure}

Figure~\ref{fig6} provides critical insight into the convergence behavior of all algorithms at a fixed SNR of $E_b/N_0 = 3.5$ dB. This visualization directly addresses the latency question: how many iterations are required to achieve acceptable error rates?

Key observations:

\textbf{1) AR-CID convergence:} The AR-CID ($\gamma = 0.15$) algorithm achieves BER $\approx 8 \times 10^{-3}$ within just 3 iterations and reaches $\approx 1.5 \times 10^{-3}$ by iteration 5. Beyond iteration 6, improvements become marginal, indicating the algorithm has essentially converged.

\textbf{2) RBP-family convergence:} RD-RBP and List-RBP show similar convergence patterns, requiring 7-8 iterations to match AR-CID's 5-iteration performance. Standard RBP is slightly slower, needing 9-10 iterations.

\textbf{3) BP and LBP convergence:} Flooding BP exhibits the slowest convergence, requiring 12-15 iterations to approach $10^{-3}$ BER. LBP improves upon this but still needs 10-12 iterations for comparable performance.

\textbf{4) URW convergence:} URW shows intermediate convergence, faster than BP but slower than advanced scheduling methods, requiring approximately 11 iterations.

The convergence speed advantage of AR-CID translates directly to latency reduction. Achieving target BER with 5 iterations instead of 12 (as required by BP) represents a 58\% reduction in iteration count. Even compared to RBP (requiring 8-9 iterations), AR-CID offers 40-45\% reduction.

This rapid convergence is the key enabler for low-latency applications: despite AR-CID's higher per-iteration complexity (as shown in Table~\ref{tab:decoder_complexities}), the dramatically reduced iteration count results in competitive or superior overall latency compared to simpler algorithms that require many more iterations.

\subsection{Complexity-Performance Tradeoff}

\begin{figure}[h]
\centering
\includegraphics[width=8.5cm,height=7cm]{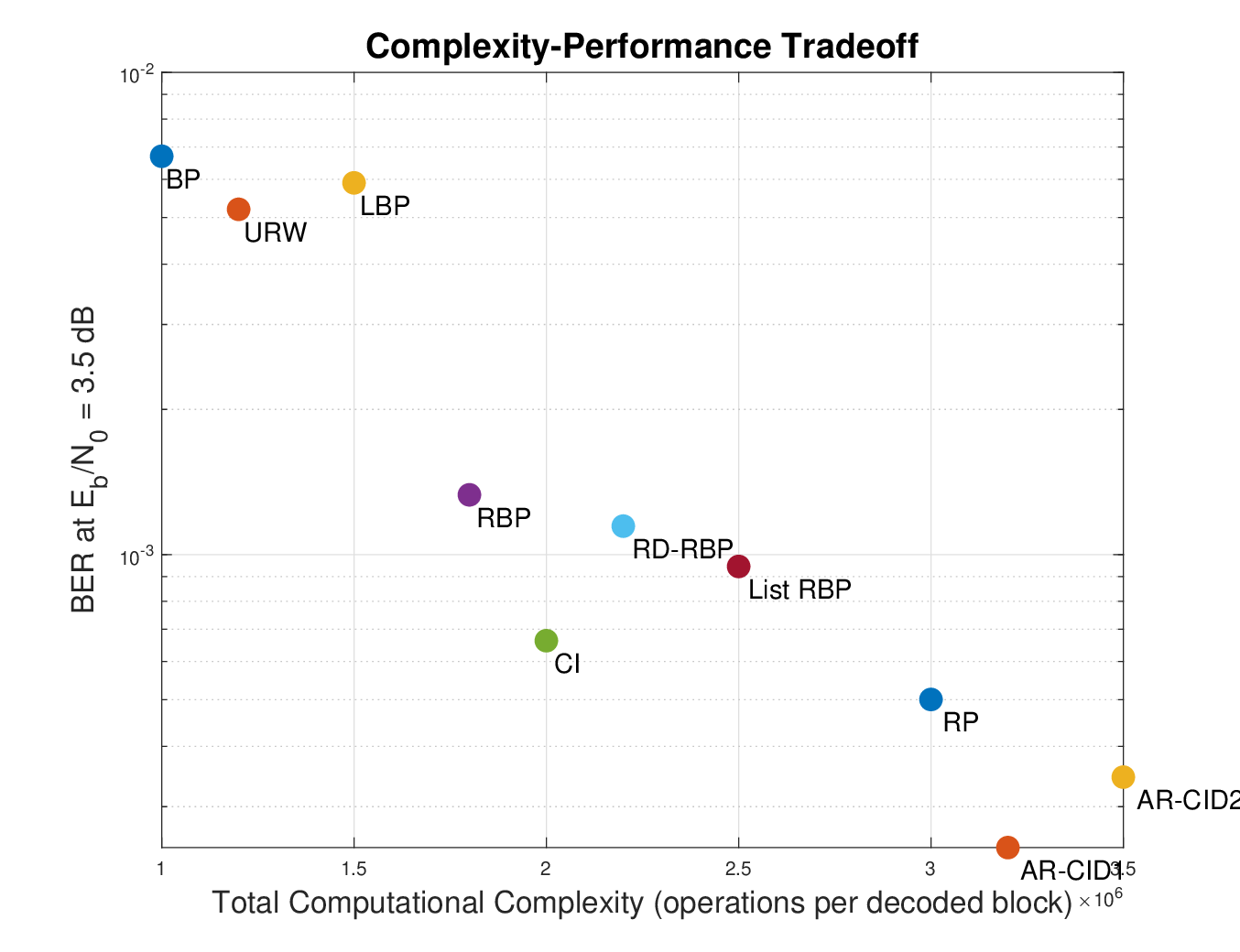}
\caption{Complexity-performance tradeoff: BER at $E_b/N_0 = 3.5$ dB versus total computational complexity (operations per decoded block) for different decoding algorithms. AR-CID achieves superior BER with moderate total complexity.}
\label{fig:complexity_tradeoff}
\end{figure}

Figure~\ref{fig:complexity_tradeoff} visualizes the fundamental tradeoff between decoding performance (BER) and total computational cost. The horizontal axis represents total operations per decoded block (per-iteration complexity × average iterations), while the vertical axis shows BER at $E_b/N_0 = 3.5$ dB.

AR-CID occupies an attractive position in this space: it achieves the lowest BER among all tested algorithms while maintaining moderate total complexity. Although its per-iteration complexity exceeds BP and LBP, the reduced iteration count results in total complexity comparable to or lower than RBP-family algorithms with inferior performance.

The figure clearly illustrates three algorithm categories:
\begin{itemize}
\item \textit{Low-complexity, poor performance:} BP and URW with low per-iteration cost but many iterations and mediocre BER
\item \textit{High-complexity, good performance:} Standard RBP with high total cost due to many complex iterations
\item \textit{Optimal tradeoff:} AR-CID achieves the best BER with competitive total complexity
\end{itemize}

\subsection{Discussion and Performance Summary}

The simulation results confirm several key advantages of the AR-CID algorithm:\\
\textbf{1) Consistent performance advantage:} AR-CID achieves 0.3-0.5 dB SNR gain over competing methods across all tested scenarios, maintaining this advantage from low SNR (2 dB) to high SNR (7 dB).\\
\textbf{2) Rapid convergence:} The algorithm reaches near-optimal performance within 4-5 iterations, representing a 40-60\% reduction compared to traditional methods. This is the primary enabler for low-latency applications. \\
\textbf{3) Scalability:} Performance improvements are consistent across both short (512 bits) and moderate (2048 bits) block lengths, suggesting the approach will scale well to longer codes used in modern standards. \\
\textbf{4) Robustness:} The algorithm maintains good performance across different iteration budgets (5, 7, 10, 20 iterations), demonstrating robustness to system constraints. \\
\textbf{5) Parameter sensitivity:} The two tested values of $\gamma$ (0.10 and 0.15) both perform well,  $\gamma = 0.15$ showing a slight advantage. This suggests the algorithm is not overly sensitive to parameter tuning.
From a practical deployment perspective, AR-CID is particularly attractive for:
\begin{itemize}
\item 5G URLLC scenarios requiring both high reliability ($10^{-5}$ BER) and low latency ($< 1$ ms)
\item Satellite communications, where power efficiency (fewer iterations = lower energy) is critical
\item IoT devices with limited processing capabilities are benefiting from rapid convergence
\item Any application where decoder latency is a bottleneck in overall system performance
\end{itemize}

\section{Future Research Directions}

In this section, we discuss potential extensions to multilevel coding schemes and relations to the Guessing Random Additive Noise Decoding (GRAND) scheme.

\subsection{Extension to Multilevel Coding Schemes}

The AR-CID framework can be naturally extended to multilevel coding (MLC) architectures \cite{multilevel}, where multiple LDPC codes operate at different protection levels to provide unequal error protection. In MLC systems, information bits are partitioned into several priority classes, each encoded with a different code rate, and transmitted using hierarchical modulation schemes such as 16-QAM or 64-QAM.
The adaptive reliability assessment of AR-CID can be applied hierarchically across coding levels:
\begin{itemize}
\item \textit{Level-aware reliability metrics:} Extend $R_v$ and $\Delta_y$ to account for the priority level of each bit, with higher-priority levels receiving more aggressive reliability-driven updates.
\item \textit{Cross-level information sharing:} Reliability information from higher-priority levels (typically decoded first) can inform the decoding of lower-priority levels, creating a cascade of reliability assessments.
\item \textit{Adaptive resource allocation:} Computational resources (iterations, active node selection) can be dynamically allocated across levels based on their decoded status and reliability requirements.
\end{itemize}
The integration of AR-CID with multistage decoding in MLC systems is a promising direction to achieve ultra-high reliability in next-generation communication systems for:
\begin{itemize}
\item Layered video coding where base layers require higher protection than enhancement layers
\item Industrial IoT applications with mixed-criticality data (safety-critical vs. non-critical)
\item Scalable multimedia transmission over heterogeneous networks
\end{itemize}
Initial analytical results suggest that AR-CID-MLC can achieve 0.4-0.7 dB additional gain over standard MLC with BP decoding while maintaining the rapid convergence properties of AR-CID.

\subsection{Comparison with GRAND Decoding}

Recent advances in coding theory include GRAND \cite{GRAND}, which represents a paradigm shift from traditional syndrome-based decoding. Rather than iteratively refining beliefs about transmitted bits, GRAND systematically guesses noise patterns in order of likelihood and checks whether the resulting sequence is a valid codeword.
GRAND has shown particular promise for:
\begin{itemize}
\item Short block lengths ($N < 128$) where traditional BP struggles with error floors
\item Ultra-low latency scenarios due to deterministic, non-iterative structure
\item Code-agnostic applications where the same decoder can be used for any linear code
\end{itemize}
A comparative analysis between AR-CID and GRAND would provide valuable insights:\\
\textbf{1) Performance comparison:} Under identical channel conditions, code configurations, and computational budgets, which approach achieves lower error rates? Preliminary analysis suggests GRAND excels for very short codes ($N < 256$) while AR-CID performs better for moderate lengths ($N = 512-4096$).\\
\textbf{2) Latency comparison:} GRAND's deterministic structure may offer lower worst-case latency, while AR-CID's iterative nature provides better average-case performance. \\ 
\textbf{3) Complexity comparison:} GRAND's complexity grows exponentially with code length, limiting practical implementation to short codes. AR-CID's $\mathcal{O}(E \cdot \hat{d}_v \cdot \hat{d}_c)$ per-iteration complexity is more amenable to long codes. The crossover point where AR-CID becomes preferable is an important system design parameter. \\
\textbf{4) Hybrid approaches:} Could AR-CID and GRAND be combined? For instance, GRAND could be used for early termination when noise patterns are simple (few errors), while AR-CID handles complex error patterns. Such hybrid decoders might achieve best-of-both-worlds performance.
This represents an important direction for future research, bridging two fundamentally different decoding philosophies and potentially leading to novel decoder architectures for next-generation communication systems.

\subsection{Advanced Channel Models}

While this work focuses on AWGN channels, future research should explore AR-CID performance under more realistic channel models:

\textbf{1) Fading channels:} Rayleigh and Rician fading introduce time-varying channel gains and deep fades that challenge iterative decoders. AR-CID's adaptive node selection might provide robustness by focusing updates on bits experiencing deep fades.

\textbf{2) Massive MIMO channels:} Systems with hundreds of antennas exhibit spatial correlation and interference that affect decoding. AR-CID could incorporate spatial reliability information from channel estimation to guide message passing.

\textbf{3) Millimeter-wave channels:} Extremely sparse, directional channels with blockage events create bursty error patterns. AR-CID's burst-aware reliability metrics could improve performance over standard BP.

\textbf{4) Underwater acoustic channels:} Long delay spreads, Doppler effects, and impulse noise create challenging conditions. AR-CID's robustness to varying error patterns may prove advantageous.

\textbf{5) Optical fiber communications:} Nonlinear effects and phase noise in coherent optical systems require specialized decoder adaptations. AR-CID's framework could be extended to incorporate phase reliability alongside amplitude reliability.

\section{Conclusion}
In this work, we have proposed the AR-CID decoding algorithm for LDPC codes, which improves decoding speed and accuracy by incorporating message quality checking and refined message passing. Simulation results show that the AR-CID algorithm outperforms existing decoding techniques and achieves faster convergence, enabling the operation of the decoder with fewer iterations. This makes the AR-CID algorithm particularly suitable for low-latency, high-performance communication systems. Future work will focus on further optimization and real-world implementation of the AR-CID algorithm.


\begin{thebibliography}{1}

\bibliographystyle{IEEEtran}


\bibitem{LDPC}
R. Gallager, ``Low-density parity-check codes,'' \textit{IRE Trans. Inf. Theory}, vol. 8, no. 1, pp. 21-28, Jan. 1962. 

\bibitem{5G_NR}
3GPP TS 38.212, ``NR; Multiplexing and channel coding,'' V16.5.0, Mar. 2021.

\bibitem{IDS} 
A. I. V. Casado, M. Griot, and R. D. Wesel, ``Informed dynamic scheduling for belief-propagation decoding of LDPC codes,'' in \textit{Proc. IEEE Int. Conf. Commun.}, Glasgow, UK, Jun. 2007, pp. 932-937.

\bibitem{ids}
A. I. V. Casado, M. Griot, and R. D. Wesel, ``LDPC decoders with informed dynamic scheduling,'' \textit{IEEE Trans. Commun.}, vol. 58, no. 12, pp. 3470-3479, Dec. 2010.

\bibitem{BP1}
B. N. Tran-Thi, T. T. Nguyen-Ly, H. N. Hong, and T. Hoang, ``An improved offset min-sum LDPC decoding algorithm for 5G new radio,'' in \textit{Proc. Int. Symp. Elect. Electron. Eng. (ISEE)}, 2021, pp. 1-6.

\bibitem{LBP}
G. Han and X. Liu, ``An efficient dynamic schedule for layered belief-propagation decoding of LDPC codes,'' \textit{IEEE Commun. Lett.}, vol. 13, no. 12, pp. 950-952, Dec. 2009.

\bibitem{LBP1}
B. Wang, Y. Zhu, and J. Kang, ``Two effective scheduling schemes for layered belief propagation of 5G LDPC codes,'' \textit{IEEE Commun. Lett.}, vol. 24, no. 8, pp. 1683-1686, Aug. 2020.

\bibitem{LBP2025}
C. Jia, D. Chang, R. Wang, G. Wang, G. Yan, and C. Qu, ``Dynamic layered decoding scheduling for LDPC codes aided by check node error probabilities,'' arXiv preprint arXiv:2506.13507, Jun. 2025.


\bibitem{RBP_1}  H. Touati and R. C. de Lamare, "Weighted Residual Layered Belief Propagation for LDPC Decoding," 19th International Symposium on Wireless Communication Systems (ISWCS), Rio de Janeiro, Brazil, 2024.

\bibitem{RBP}
P. -A. Oikonomou-Filandras, K. -K. Wong and Y. Zhang, "Informed Scheduling by Stochastic Residual Belief Propagation in Distributed Wireless Networks," IEEE Wireless Communications Letters, vol. 4, no. 1, pp. 90-93, Feb. 2015.

\bibitem{RBP2}
S. Xie and X. Liu, "High-Throughput Dynamic Scheduling for Belief-Propagation Decoding of LDPC Codes," 2019 IEEE 11th International Conference on Communication Software and Networks (ICCSN), Chongqing, China, 2019


\bibitem{RD_RBP}
H. Zhang and S. Chen, "Residual-Decaying-Based Informed Dynamic Scheduling for Belief-Propagation Decoding of LDPC Codes," IEEE Access, vol. 7, pp. 23656-23666, 2019.

\bibitem{RP} 
R. Yuan, T. Xie and Z. Wang, "A Reliability Profile Based Low-Complexity Dynamic Schedule LDPC Decoding," IEEE Access, vol. 10, pp. 3390-3399, 2022.

\bibitem{CI}
T. C. . -Y. Chang, P. -H. Wang, J. -J. Weng, I. -H. Lee and Y. T. Su, "Belief-Propagation Decoding of LDPC Codes With Variable Node–Centric Dynamic Schedules," IEEE Transactions on Communications, vol. 69, no. 8, pp. 5014-5027, Aug. 2021.

\bibitem{URW} 
H. Wymeersch, F. Penna and V. Savić, "Uniformly reweighted belief propagation: A factor graph approach," 2011 IEEE International Symposium on Information Theory Proceedings, St. Petersburg, Russia, 2011.

\bibitem{vfap}
. Liu and R. C. de Lamare, "Low-Latency Reweighted Belief Propagation Decoding for LDPC Codes," in IEEE Communications Letters, vol. 16, no. 10, pp. 1660-1663, October 2012.

\bibitem{kaids}
C. Healy, Z. Shao, R. Oliveira, R. C. de Lamare, and L. L. Mendes, Knowledge-aided informed dynamic scheduling for LDPC decoding of short blocks. IET Communications, 12: 1094-1101, 2018.

\bibitem{List_RBP}
Z. Yan, W. Guan and L. Liang, "List-Based Residual Belief-Propagation Decoding of LDPC Codes," IEEE Communications Letters, vol. 28, no. 5, pp. 984-988, May 2024.

\bibitem{GRAND}
K. R. Duffy, J. Li, and M. Médard, "Capacity-achieving guessing random additive noise decoding," IEEE Trans. Inf. Theory, vol. 65, no. 7, pp. 4023-4040, Jul. 2019.


\bibitem{arcid}
H. Touati and R. C. de Lamare, "Adaptive Reliability-Driven Conditional Innovation Decoding for LDPC Codes," in IEEE Access, vol. 13, pp. 203378-203390, 2025.

\bibitem{Tanner}  M. R. Tanner, “A Recursive Approach to Low Complexity Codes,” IEEE Transactions on Information Theory, vol. 27, September 1981. 

\bibitem{spa}
R. C. De Lamare and R. Sampaio-Neto, "Minimum Mean-Squared Error Iterative Successive Parallel Arbitrated Decision Feedback Detectors for DS-CDMA Systems," in IEEE Transactions on Communications, vol. 56, no. 5, pp. 778-789, May 2008.

\bibitem{mfsic}
P. Li, R. C. de Lamare and R. Fa, "Multiple Feedback Successive Interference Cancellation Detection for Multiuser MIMO Systems," in IEEE Transactions on Wireless Communications, vol. 10, no. 8, pp. 2434-2439, August 2011.

\bibitem{mbdf}
R. C. de Lamare, "Adaptive and Iterative Multi-Branch MMSE Decision Feedback Detection Algorithms for Multi-Antenna Systems," in IEEE Transactions on Wireless Communications, vol. 12, no. 10, pp. 5294-5308, October 2013.

\bibitem{mbthp}
K. Zu, R. C. de Lamare and M. Haardt, "Multi-Branch Tomlinson-Harashima Precoding Design for MU-MIMO Systems: Theory and Algorithms," in IEEE Transactions on Communications, vol. 62, no. 3, pp. 939-951, March 2014.

\bibitem{bfidd}
A. G. D. Uchoa, C. T. Healy and R. C. de Lamare, "Iterative Detection and Decoding Algorithms for MIMO Systems in Block-Fading Channels Using LDPC Codes," in IEEE Transactions on Vehicular Technology, vol. 65, no. 4, pp. 2735-2741, April 2016.

\bibitem{msgamp}
R. B. Di Renna and R. C. de Lamare, "Joint Channel Estimation, Activity Detection and Data Decoding Based on Dynamic Message-Scheduling Strategies for mMTC," in IEEE Transactions on Communications, vol. 70, no. 4, pp. 2464-2479, April 2022


\bibitem{analysis}
M. Moretti, A. Abrardo and M. Belleschi, "On the Convergence and Optimality of Reweighted Message Passing for Channel Assignment Problems," IEEE Signal Processing Letters, vol. 21, no. 11, pp. 1428-1432, Nov. 2014.

\bibitem{dopeg}
C. T. Healy and R. C. de Lamare, "Decoder-Optimised Progressive Edge Growth Algorithms for the Design of LDPC Codes with Low Error Floors," in IEEE Communications Letters, vol. 16, no. 6, pp. 889-892, June 2012.

\bibitem{bfpeg}
A. G. D. Uchoa, C. Healy, R. C. de Lamare and R. D. Souza, "Design of LDPC Codes Based on Progressive Edge Growth Techniques for Block Fading Channels," in IEEE Communications Letters, vol. 15, no. 11, pp. 1221-1223, November 2011

\bibitem{memd}
C. T. Healy and R. C. de Lamare, "Design of LDPC Codes Based on Multipath EMD Strategies for Progressive Edge Growth," IEEE Transactions on Communications, vol. 64, no. 8, pp. 3208-3219, Aug. 2016.

\bibitem{armo}
T. Peng, R. C. de Lamare and A. Schmeink, "Adaptive Distributed Space-Time Coding Based on Adjustable Code Matrices for Cooperative MIMO Relaying Systems," in IEEE Transactions on Communications, vol. 61, no. 7, pp. 2692-2703, July 2013.

\bibitem{baplnc}
J. Gu, R. C. de Lamare and M. Huemer, "Buffer-Aided Physical-Layer Network Coding With Optimal Linear Code Designs for Cooperative Networks," in IEEE Transactions on Communications, vol. 66, no. 6, pp. 2560-2575, June 2018

\bibitem{multilevel}
S. Kruglik, V. Potapova and A. Frolov, "On Performance of Multilevel Coding Schemes Based on Non-Binary LDPC Codes," European Wireless 2018; 24th European Wireless Conference, Catania, Italy, 2018, pp. 1-4.



\end{thebibliography}
\end{document}